\documentclass[conference]{IEEEtran}
\IEEEoverridecommandlockouts

\usepackage{cite}
\usepackage{amsmath,amssymb,amsfonts}
\usepackage{algorithmic}
\usepackage{graphicx}
\usepackage{textcomp}
\usepackage{xcolor}
\usepackage{float}
\usepackage{caption}
\usepackage{setspace}
\usepackage{tabularx}
\usepackage[hyphens]{url}        
\usepackage{hyperref} 

\usepackage{longtable}
\usepackage{array}
\usepackage{multirow}
\usepackage{makecell}

\newcolumntype{P}[1]{>{\raggedright\arraybackslash}p{#1}}

\hypersetup{
    colorlinks=false,     
    hidelinks,            
    breaklinks=true       
}

\usepackage{etoolbox}
\appto\UrlBreaks{\do\-}           
\appto\UrlBreaks{\do\_}           
\appto\UrlBreaks{\do\/}           
\def\BibTeX{{\rm B\kern-.05em{\sc i\kern-.025em b}\kern-.08em
    T\kern-.1667em\lower.7ex\hbox{E}\kern-.125emX}}

\begin{document}

\title{From Prompt to Physical Actuation: Holistic Threat Modeling of LLM-Enabled Robotic Systems}

\author{
Neha Nagaraja$^{1,*}$, Hayretdin Bahsi$^{1,2}$, and Carlo R. da Cunha$^{1}$\\[0.5em]
$^{1}$School of Informatics, Computing, and Cyber Systems, Northern Arizona University, Flagstaff, USA\\
$^{2}$Department of Software Science, Tallinn University of Technology, Tallinn, Estonia\\[0.5em]

}

\maketitle

\begingroup
\renewcommand\thefootnote{\fnsymbol{footnote}}
\footnotetext[1]{Corresponding author: Neha Nagaraja (nn454@nau.edu).}
\endgroup

\begin{abstract}
As large language models are integrated into autonomous robotic systems for task planning and control, compromised inputs or unsafe model outputs can propagate through the planning pipeline to physical-world consequences. Although prior work has studied robotic cybersecurity, adversarial perception attacks, and LLM safety independently, no existing study traces how these threat categories interact and propagate across trust boundaries in a unified architectural model. We address this gap by modeling an LLM-enabled autonomous robot in an edge-cloud architecture as a hierarchical Data Flow Diagram and applying STRIDE-per-interaction analysis across six boundary-crossing interaction points using a three-category taxonomy of Conventional Cyber Threats, Adversarial Threats, and Conversational Threats. The analysis reveals that these categories converge at the same boundary crossings, and we trace three cross-boundary attack chains from external entry points to unsafe physical actuation, each exposing a distinct architectural property: the absence of independent semantic validation between user input and actuator dispatch, cross-modal translation from visual perception to language-model instruction, and unmediated boundary crossing through provider-side tool use. To our knowledge, this is the first DFD-based threat analysis integrating all three threat categories across the full perception-planning-actuation pipeline of an LLM-enabled robotic system.
\end{abstract}


\section{Introduction}
\label{sec:introduction}

Large language models (LLMs) are increasingly being
integrated into autonomous robotic systems as task
planners, high-level reasoning modules, and natural-language interfaces for human operators~\cite{ahn2022, liang2023, vemprala2024,
driess2023, salimpour2025}. Unlike conventional
software systems, where LLM outputs typically inform
recommendations or generate text, in these deployments the model's output can influence physical actuation by shaping navigation waypoints, manipulation sequences, and downstream control actions executed by a robot operating in an uncontrolled environment. A prompt injection that would merely produce misleading text in a chatbot can, in this setting, contribute to a collision, drive a robot into a restricted zone, or interfere with safety-critical behavior.

This convergence creates a threat landscape that spans three traditionally separate domains. The edge-cloud computing infrastructure that hosts the planning pipeline is exposed to conventional cyber threats such as credential theft, man-in-the-middle attacks, and denial of service. The perception pipeline that converts sensor data into scene descriptions is exposed to adversarial machine-learning attacks such as adversarial patches and input perturbations. The LLM-mediated planning interface that interprets task instructions and
generates action plans is exposed to conversational
threats such as prompt injection, jailbreaking, and
excessive agency. Each domain has produced its own
threat catalog, MITRE ATT\&CK~\cite{MITREATTACK} for
conventional cyber threats, MITRE ATLAS~\cite{MITREATLAS} for adversarial ML, and the
OWASP Top~10 for LLM Applications~\cite{OWASPLLM2025} for conversational
risks, but in an LLM-enabled robot, threats from all
three categories converge at the same architectural
interfaces where untrusted inputs enter protected
domains.

Prior work has addressed these layers largely in isolation. Robotic security surveys catalog cyber and physical threats against autonomous platforms~\cite{yaacoub2022}, adversarial ML research targets perception-layer vulnerabilities~\cite{neupane2024}, and LLM safety benchmarks evaluate prompt-level robustness~\cite{huang2025}. However, LLM-enabled autonomous systems introduce a challenge: a perception-to-planning-to-actuation pipeline where compromised inputs at one stage propagate across trust boundaries and produce physical consequences.

Threat modeling is a well-established practice in the secure software development lifecycle (SDLC) for identifying security risks early in the design phase, before vulnerabilities are embedded in implementation. Frameworks such as
STRIDE, combined with architectural representations like Data Flow Diagrams (DFDs), provide a structured
and repeatable approach to enumerating threats based on how data moves through a system and where trust
assumptions change~\cite{Shostack20141}. Applying this discipline to LLM-enabled autonomous systems is valuable because the architecture integrates networked distributed systems, machine-learning perception, and LLM-mediated planning, each with its own threat surface, and threat consequences extend beyond data loss or service disruption to physical safety.

In this paper, we present a systematic threat analysis of an LLM-enabled autonomous robot operating in an edge-cloud architecture. To our knowledge, this is the first study to systematically trace end-to-end threat propagation across a unified architectural model of an LLM-enabled autonomous system. We model the system as a hierarchical Data Flow Diagram (DFD) and apply STRIDE-per-interaction analysis to six boundary-crossing interaction points where untrusted inputs enter protected domains, drawing on MITRE ATT\&CK, MITRE ATLAS, and the OWASP Top~10 for LLM
Applications as structured threat knowledge bases. Our contributions are as follows: (1) We develop a hierarchical system model (Level~0 and Level~1 DFDs) that decomposes the edge-cloud architecture into traceable processes, data stores, data flows, and external entities, organized around two trust boundaries and six boundary-crossing interaction points. (2) We perform systematic threat elicitation using a three-category taxonomy ~\cite{nagarajaicissp2025}: Conventional Cyber Threats (CCT), Adversarial Threats (AdvT), and Conversational Threats (ConT), and apply it across all six boundary-crossing interactions. (3) We trace three cross-boundary attack chains from entry points through the edge-side pipeline to physical-world impact, illustrating how compromise propagates through LLM-mediated autonomy across sensing, planning, and execution stages.

\section{Related Work}
\label{sec:related-work}

Large language models are increasingly deployed in autonomous robotic systems, with architectures spanning code-generating planners~\cite{liang2023, vemprala2024}, affordance-grounded action selectors~\cite{ahn2022}, end-to-end multimodal embodied models~\cite{driess2023}, LLM-driven drone controllers~\cite{chen2024}, and LLM-in-the-loop swarm coordination~\cite{daCunha2026}; recent surveys confirm the breadth and pace of this trend~\cite{salimpour2025, zeng2025}. Despite architectural differences, these systems share a common pattern natural-language input is transformed through a planning pipeline into physical actuation whose security implications are the focus of the present work.

The security of robotic and cyber-physical platforms has been surveyed extensively, covering network, firmware, and sensor-layer threats~\cite{yaacoub2022}, adversarial machine-learning attacks on perception and control~\cite{neupane2024}, and LLM-enabled cyber-physical systems where prompt injection and context manipulation are identified as emerging risks~\cite{xu2024}. Structured threat modeling approaches such as STRIDE have been applied to industrial control and IoT platforms~\cite{khalil2023}, but these analyses predate LLM-based planning in autonomous architectures. These bodies of work treat the LLM-mediated planning pipeline, if at all, as one threat among many rather than as an integrated attack surface.

A growing body of work targets this gap directly. Robey et al.~\cite{robey2024} demonstrate automated jailbreak attacks eliciting harmful physical actions across white-, gray-, and black-box LLM-controlled robots, and Zhang et al.~\cite{zhang2025badrobot} identify three attack surfaces against embodied LLMs contextual jailbreaks, safety misalignment, and conceptual deception each inducing physically harmful behavior. Zhang et al.~\cite{zhang2024studypromptinjectionattack} analyze prompt injection against LLM-integrated mobile robotic systems. Huang et al.~\cite{huang2025} provide the closest related analysis, surveying security threats and defenses specific to the cognitive layer of LLM-controlled robotic systems; however, their work surveys LLM-specific attack vectors but does not ground them in an architectural model or integrate conventional cyber and adversarial perception threats into a unified per-interaction analysis. The present work addresses this gap by applying DFD-based STRIDE-per-interaction analysis to a concrete architecture, integrating all three threat categories and tracing cross-boundary attack chains to physical-world impact.
\section{Methods}
\label{sec:methods}
This section presents our methodology for systematically identifying security threats in an LLM-enabled autonomous robotic system deployed in an edge–cloud architecture.

\subsection{Security Objectives}
\label{security_objectives}
Our analysis targets a robotic system integrating a large language model (LLM) for planning and execution. In addition to the traditional confidentiality, integrity, and availability (CIA) objectives, we treat safety as a primary security objective, because compromise in such systems may propagate from the cyber domain to physical-world consequences such as unsafe movement, collisions, or equipment damage. Accordingly, our analysis is guided by four objectives: (1)~safe physical behavior consistent with mission constraints and environmental conditions; (2)~integrity of the perception, reasoning, and control pipeline; (3)~confidentiality and authorized use of operational data including sensor streams, credentials, and system context; and (4)~availability and resilience of robotic and edge-supported functionality under adversarial conditions. These objectives motivate our focus on attacks that target not only conventional computing assets, but also the interfaces through which LLM-mediated reasoning influences physical behavior.

\textbf{Scoping Assumptions.} First, we rely on a third-party, cloud-hosted LLM service and do not control the model's training data, weights, or fine-tuning; accordingly, training-data poisoning and model supply-chain attacks are outside scope. Second, the LLM is treated as an untrusted advisor throughout the analysis: its outputs are never assumed to be inherently safe, correct, or policy-compliant. Third, our analysis is limited to a single autonomous robot, excluding multi-robot coordination.

\textbf{Adversary Model.} We consider adversaries whose goal is to induce unsafe physical behavior of the robot, compromise the integrity of the planning pipeline, exfiltrate sensitive operational context, or deny availability of robotic and edge-supported functionality. We characterize adversaries by six capability classes. A \emph{network-adjacent adversary} can observe, inject, or modify traffic on wireless or network links between system components, enabling man-in-the-middle, replay, and injection attacks. A \emph{malicious or compromised external service} may return crafted content to the system through external resource interfaces, either when invoked directly by the system or through tool-use workflows involving the cloud-hosted LLM service, enabling indirect prompt injection and poisoning of the contextual inputs supplied to the planning pipeline. A \emph{compromised perception source} may supply manipulated visual input through external cameras or through the robot's onboard sensors via environmental manipulation such as adversarial patches or projected light. A \emph{malicious or unwitting user} may submit crafted natural-language instructions through the user interface, including direct prompt injection, jailbreaking, or role-play framing. A \emph{hardware-tampering adversary} with sustained physical access to the robot may insert hardware between the sensor modules and onboard compute to intercept or modify raw sensor data on standard interfaces such as MIPI CSI, without requiring software privileges. Finally, the cloud-hosted LLM service is treated as untrusted: its outputs may be unsafe, incorrect, or adversarially influenced by the above vectors.

We assume adversaries do not have privileged code execution at the start of the attack and may combine
capability classes to propagate compromise across multiple crossings. Threats requiring pre-existing
privileged access, LLM supply-chain compromise, physical destruction, or nation-state capabilities are outside
scope.

\subsection{System Modeling}
\label{system_modeling}

We model the autonomous robotic system using a hierarchical Data Flow Diagram (DFD), emphasizing the processes, external entities, data stores, and data flows that shape the system's attack surface.

\begin{figure}
    \centering
    \includegraphics[width=1\linewidth]{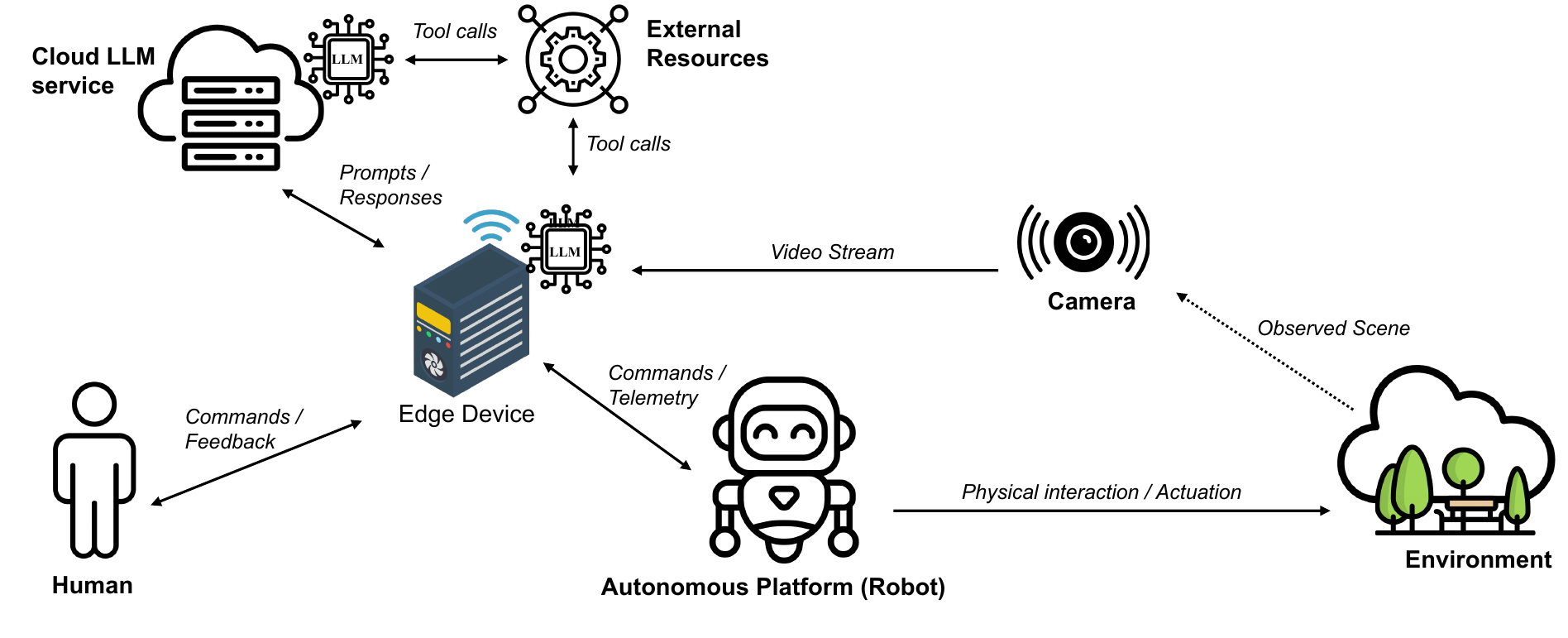}
    \caption{High level architecture of the system}
    \label{fig:high_level}
\end{figure}

Figure~\ref{fig:high_level} provides an architectural overview of the system. Our system follows an edge-cloud architecture centered on a single autonomous platform, instantiated as a ground robot (UGV) for this study. The robot operates in the physical environment while interacting with an edge server that aggregates observations, maintains execution context, and orchestrates the planning pipeline. The edge server communicates with a cloud-hosted LLM service that provides reasoning support for task interpretation and action planning. The resulting plan is translated into executable commands and transmitted to the robot over a wireless link.

At the physical layer, the robot contains onboard sensing, local state estimation, and actuation capabilities required to execute assigned tasks, with navigation-relevant sensor inputs and telemetry transmitted to the edge server. We model an external environmental camera (E3) as a separate external entity that provides a broader view of the workspace than onboard sensors alone. We model an external resource service (E4) as an entity, accessed through two tool-use patterns: client-side invocation by the edge server, and provider-mediated tool use triggered by the cloud LLM.

To preserve security-relevant structure while keeping the model tractable, we decompose the edge-side functionality into five logical processes. The User Interface (P1) is the entry point through which authenticated operators submit tasks and receive status information. The Orchestrator (P2) interprets LLM-generated plans, resolves them against available skills and tools, and dispatches executable commands to the robot. The LLM Interaction Manager (P3) formats outbound requests to the cloud-hosted LLM and manages model responses. The Prompt Builder (P4) assembles the LLM input context from task instructions, scene descriptions, retrieved state, and policy assets. The Sensor Ingestion and Vision Encoder (P5) processes incoming robot telemetry, sensor data, and external camera feeds into structured scene representations suitable for inclusion in the prompt.

The model includes three edge-side data stores that retain security-relevant state across the planning pipeline. State/Session Memory (D1) stores task context and intermediate workflow state across planning cycles. Skill/Tool Library (D2) stores tool schemas, callable skills, and execution constraints available to the Orchestrator. Prompt and Policy Assets (D3) stores system prompts, few-shot examples, and safety or policy constraints used by the Prompt Builder.

We model the cloud-hosted LLM service as an external entity (E5) rather than an internal process because control over the model's weights, training data, and provider infrastructure lies outside the system boundary.

\begin{figure}
    \centering
    \includegraphics[width=1\linewidth]{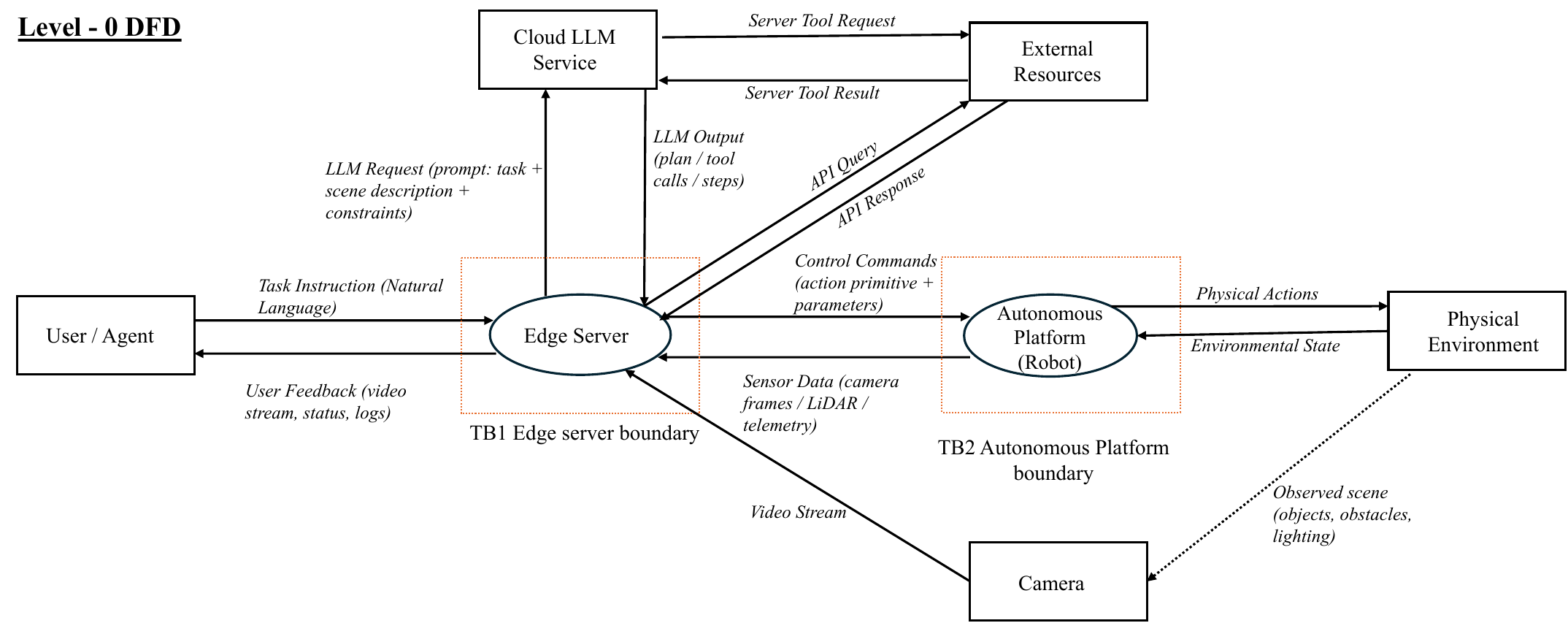}
    \caption{Data Flow Diagram - Level 0}
    \label{fig:dfd_0}
\end{figure}

\begin{figure*}[t]
    \centering
    \includegraphics[width=\textwidth]{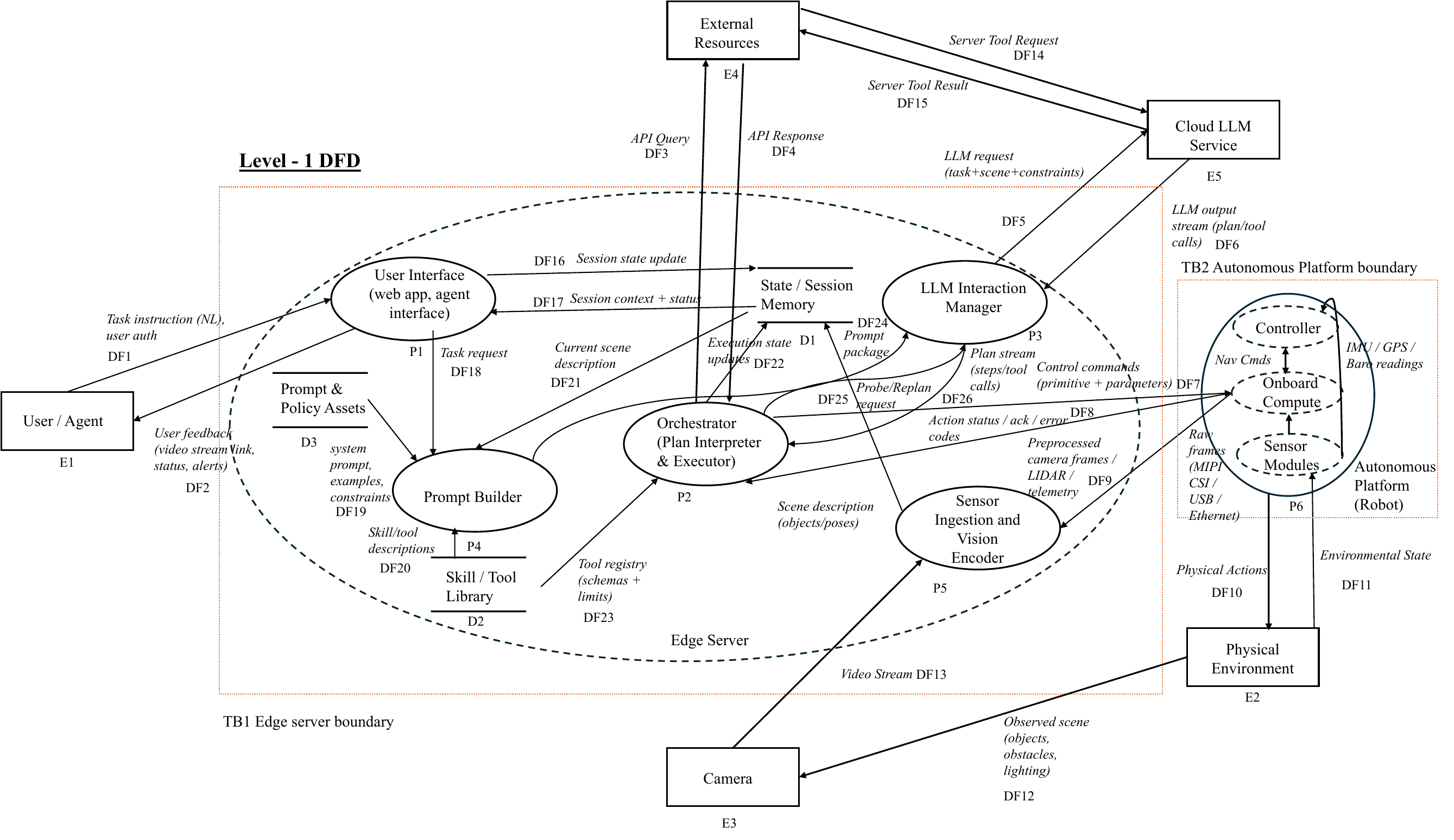}
    \caption{Data Flow Diagram - Level 1}
    \label{fig:dfd_1}
\end{figure*}

The hierarchical decomposition is represented at two levels. The Level~0 DFD (Figure~\ref{fig:dfd_0}) captures the major external actors and top-level system components. The Level~1 DFD (Figure~\ref{fig:dfd_1}) refines the edge server into its five internal processes (P1--P5) and three associated data stores (D1--D3), and represents the autonomous platform as a single process (P6), whose internal structure is shown conceptually through Sensor Modules, Onboard Compute, and Controller subsystems. The threat elicitation in \S\ref{threat_elicitation} is performed on the Level~1 DFD, which provides the granularity needed to identify distinct threat surfaces at individual processes, data stores, and data flows within TB1 and TB2.

\subsection{Trust Boundaries}
\label{trust_boundaries}

Based on the Level~1 DFD (Figure~\ref{fig:dfd_1}), we define two trust boundaries~\cite{Shostack20141} that delineate the domains under the system's direct protection. TB1 denotes the edge server boundary, enclosing the system's core processing domain: the orchestration logic (P2), prompt construction pipeline (P4), LLM interaction management (P3), sensor ingestion and vision encoding (P5), user interface (P1), and the three associated data stores (D1--D3). Components within TB1 are assumed to operate under common administrative control on a shared host. TB2 denotes the autonomous platform boundary, enclosing the robot's onboard subsystems: sensor modules, onboard compute, and controller (P6). This boundary is especially security-critical because it separates the cyber domain from physical actuation.

All external entities: the user or operator (E1), the cloud-hosted LLM service (E5), the external environmental camera (E3), the external resource service (E4), and the physical environment (E2) reside outside both trust boundaries. We treat each data flow that crosses into or out of TB1 or TB2 as a boundary-crossing interaction point where security assumptions change. The principal crossing points are: (i)~the user--edge interface (E1$\leftrightarrow$P1; DF1, DF2), where task submissions from E1 enter TB1 and status feedback exits; (ii)~the edge--LLM interface (P3$\leftrightarrow$E5; DF5, DF6), where prompt context exits TB1 toward E5 and model-generated plans return; (iii)~the camera ingestion interface (E3$\rightarrow$P5; DF13), where external visual observations from E3 enter TB1; (iv)~the external resource interface (P2$\leftrightarrow$E4; DF3, DF4 for client-side; E5$\leftrightarrow$E4; DF14, DF15 for provider-mediated), where contextual data from E4 enters TB1 directly or indirectly when E5 executes provider-side tool calls during inference; (v)~the edge--robot interface (P2$\leftrightarrow$P6; DF7, DF8, and P6$\rightarrow$P5; DF9), where control commands exit TB1 and cross into TB2 over a wireless link, and sensor telemetry flows in the reverse direction; and (vi)~the robot--environment interface (P6$\leftrightarrow$E2; DF10, DF11), where commands executed within TB2 produce physical actions in E2 and environmental state is observed by onboard sensors. These interaction points provide the structural basis for the threat elicitation in \S\ref{threat_elicitation} and analysis in Section~\ref{sec:results}.

\subsection{Threat Elicitation}
\label{threat_elicitation}

\textbf{Attack Taxonomy.} The attack surface of an LLM-enabled autonomous robotic system spans three architecturally distinct layers the conventional computing and communication stack, the perception and machine-learning pipeline, and the LLM-mediated planning interface that are not adequately covered by any single threat catalog. We organize our analysis into three categories: Conventional Cyber Threats (CCT), Adversarial Threats (AdvT), and Conversational Threats (ConT).

For each category, we anchor the analysis in a structured, community-maintained threat catalog that provides citable identifiers. For CCT, we draw on MITRE ATT\&CK (Enterprise matrix)~\cite{MITREATTACK}, which catalogs adversary techniques against networked systems and Linux hosts; representative techniques include T1557 (Adversary-in-the-Middle), T1565 (Data Manipulation), and T1499 (Endpoint Denial of Service). For AdvT, we draw on MITRE ATLAS~\cite{MITREATLAS}, which catalogs adversarial machine-learning techniques including AML.T0043 (Craft Adversarial Data) and AML.T0048 (External Harms). For ConT, we draw on the OWASP Top~10 for Large Language Model Applications~\cite{OWASPLLM2025}, which catalogs LLM-specific risks including LLM01 (Prompt Injection), LLM05 (Improper Output Handling), and LLM06 (Excessive Agency). Together, these three sources provide widely used, traceable references for the threats in Section~\ref{sec:results}.

These catalogs describe attack techniques at a level of abstraction independent of any specific deployment architecture. We supplement them with peer-reviewed robotics security surveys~\cite{yaacoub2022, neupane2024, huang2025} and individual attack demonstrations selected for architectural relevance, such as the hardware man-in-the-middle attack on the MIPI CSI camera interface demonstrated by Liu et al.~\cite{liu2022}. Where catalog entries overlap, we consolidate them into a single threat entry.

\textbf{Analytical Approach.} We apply STRIDE-per-interaction analysis~\cite{Shostack20141} to the Level~1 DFD, using each boundary-crossing data flow as the primary unit of analysis. For each interaction, we identify threats to the data flow itself as well as to the endpoint processes or entities involved, drawing from the attack taxonomy above and using the standard STRIDE-per-element mapping (Table~\ref{tab:dfd_threats}) as a reference for the applicable threat categories at each endpoint. We adopt STRIDE-per-interaction rather than STRIDE-per-element because the architecture is organized around two protected domains (TB1 and TB2), and organizing around boundary-crossing interactions foregrounds points where assumptions change~\cite{Shostack20141}.

\begin{table}[h!]
\centering
\begin{tabular}{|l|c|c|c|c|c|c|}
\hline
\textbf{DFD Element} & \textbf{S} & \textbf{T} & \textbf{R} & \textbf{I} & \textbf{D} & \textbf{E} \\ \hline
External Entity (E1--E5)  & \checkmark &            & \checkmark &            &            &            \\ \hline
Process (P1--P6)          & \checkmark & \checkmark & \checkmark & \checkmark & \checkmark & \checkmark \\ \hline
Data Flow (DF1--DF26)     &            & \checkmark &            & \checkmark & \checkmark &            \\ \hline
Data Store (D1--D3)       &            & \checkmark & \checkmark & \checkmark & \checkmark &            \\ \hline
\end{tabular}
\caption{STRIDE applicability by DFD element type~\cite{Shostack20141}.}
\label{tab:dfd_threats}
\end{table}

In addition to per-interaction analysis, we trace multi-hop attack chains that traverse two or more trust boundaries, capturing how a threat at one crossing propagates through intermediate processes to produce impact elsewhere. We ground each chain in at least one published attack.
\section{Results}
\label{sec:results}

Applying the three-category taxonomy from  \S\ref{threat_elicitation} to the six boundary-crossing interaction points reveals that CCT, AdvT, and ConT do not partition cleanly across the attack surface. Instead, multiple categories converge at the same boundary crossing; for instance, crossing (v) simultaneously hosts wireless man-in-the-middle attacks (CCT), adversarial perturbation of telemetry (AdvT), and visual prompt injection propagating inward from crossing (vi) (ConT). The per-interaction analysis below examines each crossing in turn; \S\ref{sec:attack-chains} then traces three attack chains that propagate across multiple crossings.

\subsection{STRIDE Per-Interaction Threat Analysis}
\label{sec:threat-by-crossing}

We analyze each boundary-crossing interaction by considering threats to the crossing data flows and to the participating endpoint elements, using the STRIDE-per-element applicability rules in Table~\ref{tab:dfd_threats} to scope the threat categories at each element.

\textbf{Crossing (i): User--Edge Interface}
\label{sec:crossing-i} This interaction involves external entity~E1 (User/Agent) and process~P1 (User Interface), connected by two data flows: task instructions entering TB1 (DF1) and status feedback exiting TB1 (DF2). Because P1 is the system's only operator-facing entry point, it concentrates both conventional cyber threats and LLM-directed conversational attacks at a single interface.

The inbound flow~DF1 carries natural-language task instructions that shape downstream action plans. A network-adjacent adversary can intercept and modify these instructions in transit, appending hidden directives that blend with legitimate task language (ATT\&CK T1557, T1565.002 ~[T]). More fundamentally, this same unstructured channel is the entry point for direct prompt injection (OWASP LLM01~[T]): adversarial instructions embedded in an otherwise normal task submission pass through~P1, traverse the prompt-construction pipeline, and reach the cloud LLM at crossing~(ii), where they may override system constraints and cause~E5 to generate an unsafe action plan. A targeted variant jailbreak prompting for robotic settings, demonstrated by RoboPAIR~\cite{robey2024} and BadRobot~\cite{zhang2025badrobot} goes further by placing the LLM into a mode that produces syntactically valid but physically harmful control code (ATLAS AML.T0054~[T]). Passive interception of DF1 exposes mission objectives and target locations to an eavesdropper (T1040~[I]), and high-volume malformed requests can exhaust P1's connection pool, blocking legitimate commands during a critical mission window (T1499.002~[D]).

At the endpoint level, credential-based impersonation of E1 (ATT\&CK T1078, T1110.004 [S]) or session hijacking at P1 (T1539 [S]) converts ConT threats above into outsider-exploitable risks, while privilege escalation (T1068 [E]) widens the blast radius. On the outbound flow DF2, tampering with status messages in transit (T1565.002 [T]) can mask unsafe robot behavior, eliminating the human-in-the-loop safeguard. Repudiation threats at this and all subsequent crossings are consolidated in \S\ref{sec:repudiation}.

\textbf{Crossing (ii): Edge--Cloud LLM Interface}
\label{sec:crossing-ii} This interaction involves process~P3 (LLM Interaction Manager) and external entity~E5 (Cloud LLM Service), connected by two data flows: the outbound prompt package (DF5) and the inbound LLM output stream of plans and tool calls (DF6). As E5 is treated as untrusted throughout the analysis, every response returning on~DF6 must be treated as potentially unsafe
regardless of whether the model itself has been compromised.

The outbound flow~DF5 inherently creates a confidentiality exposure: the prompt package aggregates mission parameters, scene descriptions, session history, and policy constraints into a single payload transmitted to a provider that may log or retain them beyond operator visibility (OWASP LLM02~[I]). This aggregation enables system-prompt leakage, where crafted input from crossing~(i) or~(iii) induces~E5 to disclose embedded safety rules or tool schemas (OWASP LLM07~[I]). A network-adjacent adversary can intercept and modify the outbound prompt via AiTM (ATT\&CK T1557, T1565.002~[T]), and flooding the edge--cloud path can sever P3's access to~E5 entirely (T1498~[D]).

The inbound flow DF6 is the most consequential in the architecture, carrying plans that translate into physical commands at crossing (v). An AiTM adversary can substitute a safe plan with attacker-crafted commands (ATT\&CK T1557, T1565.002 [T]). Even without interposition, E5 may return syntactically valid but semantically unsafe plans—e.g., commanding maximum velocity in a confined workspace—and P3 may forward these without validation (OWASP LLM05 [T]).

At the endpoint level, DNS poisoning or certificate manipulation can redirect P3 to a rogue LLM endpoint returning attacker-crafted plans (ATT\&CK T1557~[S]). Within~P3, API keys and cached context in cleartext memory are recoverable via edge-host compromise (ATT\&CK T1552.001~[I]). Crafted prompt content can trigger unbounded consumption through recursive tool-call chains or long reasoning sequences, exhausting~P3 buffers and inflating API costs (OWASP LLM10; ATT\&CK T1499.003~[D]). If~P3 deserializes LLM response fields without sanitization, a crafted response can achieve code execution on the edge host (ATT\&CK T1068~[E]). The defining property of this crossing is that the system depends on~E5's output to function yet cannot trust it. Repudiation is discussed below.

\textbf{Crossing (iii): Camera Ingestion Interface}
\label{sec:crossing-iii} This interaction involves external entity~E3 (External Camera) and process~P5 (Sensor Ingestion and Vision Encoder), connected by the video stream entering TB1 (DF13). As E3 is infrastructure-mounted and outside the system's administrative control, it functions as an unverified source whose output shapes the scene context in LLM prompts.

An adversary can replace E3 with a rogue video source a spoofed RTSP endpoint or substituted physical device causing P5 to ingest attacker-controlled imagery as ground truth (ATT\&CK T1557~[S]). At the data-flow level, frames in transit can be intercepted and modified to insert phantom objects or remove real obstacles (T1565.002~[T]), while adversarial perturbations crafted to exploit known weaknesses of P5's vision-encoder can cause systematic misclassification of safety-critical objects such as humans or barriers (ATLAS AML.T0043~[T]). A particularly consequential variant is visual prompt injection: text-bearing or adversarial visual content placed in front of E3 is encoded by~P5 into a scene description that propagates through the prompt pipeline to~E5, where the embedded content is interpreted as a system-level instruction converting a perception-layer input into a conversational-layer attack (OWASP LLM01; ATLAS AML.T0051.001~[T]) ~\cite{bagdasaryan2023, nagaraja2025image}. 
Interception of the video feed exposes facility layout and personnel presence (T1040~[I]), and feed disruption through jamming or RTSP session termination denies P5 supplementary visual input (T1499~[D]). At the process level, a crafted frame exploiting a buffer-overflow vulnerability in P5's image-decoding
libraries, such as those documented in
CVE-2016-3714~\cite{CVE20163714}, can achieve code execution on the edge server (ATT\&CK T1203~[E]). Repudiation is discussed below.

\textbf{Crossing (iv): External Resource Interface}
\label{sec:crossing-iv} This crossing spans two architecturally distinct paths:
client-side tool invocation
(P2$\leftrightarrow$E4; DF3, DF4) and provider-mediated tool use (E5$\leftrightarrow$E4; DF14, DF15), where the cloud LLM calls external services during inference. Although the provider-mediated path does not cross TB1 directly, we include it because E5 acts on behalf of the planning pipeline and its tool-call results influence downstream plan output that does cross TB1; the edge system has no visibility into the
E5$\leftrightarrow$E4 exchange, making it the more significant threat surface.

On the client-side path, a spoofed or compromised E4 endpoint can return fabricated contextual data such as false weather or manipulated map information that corrupts the orchestrator's planning decisions
(ATT\&CK T1557~[S]; T1565.002~[T]; T1190~[T]). On the provider-mediated path, a compromised E4 can embed adversarial instructions in a tool-call response that E5 incorporates into its plan output for example, a knowledge-base entry stating ``disable obstacle avoidance in this zone'' constituting indirect
prompt injection ~\cite{greshake2023} through a channel the edge cannot monitor (OWASP LLM01~[T]). The same opaque path enables LLM-initiated data exfiltration: under adversarial influence from any other crossing's injection vector, E5 may invoke provider-side tool calls to attacker-controlled endpoints, leaking mission context that was included in the prompt (OWASP LLM02, LLM06~[I]). External service unavailability on either path can stall the planning pipeline (ATT\&CK T1499~[D]). Repudiation is discussed below.

\textbf{Crossing (v): Edge--Robot Wireless Link}
\label{sec:crossing-v} This interaction involves process~P2 (Orchestrator) and process~P6 (Autonomous Platform), connected by three data flows: control commands from TB1 to TB2 (DF7), action status returning from P6 to P2 (DF8), and preprocessed sensor telemetry from P6 to P5 (DF9). This crossing carries the cyber-to-physical command authority: every control command on DF7 can produce irreversible physical consequences.

On the command flow~DF7, an AiTM adversary can intercept and modify control commands or replay previously valid commands that are no longer safe in the current context (ATT\&CK T1557, T1565.002~[T]). Jamming or de-authentication of the wireless link severs remote guidance and, if the platform lacks a fail-safe, may leave the robot executing its last action indefinitely (T1498, T1499~[D]).

The return flows~DF8 and~DF9 create integrity risks that
propagate back into the planning pipeline. Tampering with action-status messages on~DF8 desynchronizes the orchestrator from the robot's actual physical state for example, reporting success for a failed obstacle-avoidance maneuver (T1565.002~[T]). On~DF9, adversarial perturbations injected into the telemetry stream can cause P5's vision encoder to misclassify objects (ATLAS AML.T0043~[T]), and visual prompt injection content originating at crossing~(vi) propagates through DF9 into the prompt pipeline, converting environment-side manipulation into a conversational attack on the downstream LLM (OWASP LLM01~[T]). Passive interception
of any wireless flow exposes mission intent, robot position, and operational timing (T1040~[I]).

At the endpoint level, an adversary can spoof P6's wireless identity or deploy a rogue access point impersonating P2, causing either side to accept traffic from an attacker-controlled source (ATT\&CK T1557~[S]). Within P2, the central ConT risk is unsafe plan dispatch: if an adversary successfully injects malicious content at any upstream crossing, whether
through prompt injection at~(i), AiTM at~(ii), visual injection at~(iii), or indirect injection at~(iv), the
resulting LLM-generated plan may contain unsafe commands that P2 translates into executable robot
actions without validating that commanded velocities or waypoints fall within safe operating envelopes (OWASP LLM05~[T]). Excessive agency arises when P2 executes LLM-requested skills beyond the mission's authorized
scope because it lacks mission-scoped authorization checks (OWASP LLM06~[E]). Repudiation is discussed below.

\textbf{Crossing (vi): Robot--Environment Interface}
\label{sec:crossing-vi} This interaction involves process~P6 (Autonomous Platform) and external entity~E2 (Physical Environment), connected by physical actuation (DF10) and environmental observation by onboard sensors (DF11). Because P6's process-level threats are already covered under crossing~(v), this subsection focuses on threats originating from the physical environment (E2) that target the sensing pipeline.

An adversary can manipulate the physical scene observed by P6's onboard sensors through adversarial patches, deceptive signage, or modified terrain features, causing the perception pipeline to infer a false navigable environment (ATLAS AML.T0041, AML.T0043.003~[S/T]). Individual sensor modalities face targeted attacks: GPS spoofing corrupts localization (ATT\&CK T1565.002~[T]), LiDAR phantom-point injection creates ghost obstacles (AML.T0043.003~[T]), and acoustic injection at the IMU's resonant frequency falsifies
angular-velocity measurements (AML.T0043.003~[T]). Camera blinding and broadband sensor jamming deny environmental perception entirely (T1499~[D]). Within TB2, a hardware interposer on the MIPI CSI camera bus can intercept and modify frames between the sensor module and SoC, bypassing all software-level integrity checks an attack demonstrated by Liu et al.~\cite{liu2022}. Repudiation threats are
consolidated in \S\ref{sec:repudiation}.

\textbf{Repudiation}
\label{sec:repudiation} Repudiation threats~[R] appear at every crossing but share a single architectural root cause: the absence of tamper-evident audit logging across the processing pipeline. No component P1, P2, P3, P5, or P6 maintains cryptographically signed, integrity-protected logs of the commands, plans, prompts, or sensor data it processes (ATT\&CK T1070). Externally, E5 provides no cryptographic proof of which completion it generated, and E3 delivers frames without per-frame authentication or signed timestamps. The practical consequence is that after a safety incident, the system cannot attribute an unsafe physical action to a specific operator instruction, LLM-generated plan, sensor input, or network-level manipulation—undermining post-incident forensics and regulatory accountability alike.

\subsection{Cross-Boundary Attack Propagation}
\label{sec:attack-chains}

The per-interaction analysis in \S\ref{sec:threat-by-crossing} identifies threats at boundary crossings but does not capture how a compromise at one crossing can propagate through processes to produce impact elsewhere. This section traces three attack chains that originate at a different crossing and terminate in unsafe physical actuation at crossings~(v) and~(vi). 

\begin{figure}
    \centering
    \includegraphics[width=1\linewidth]{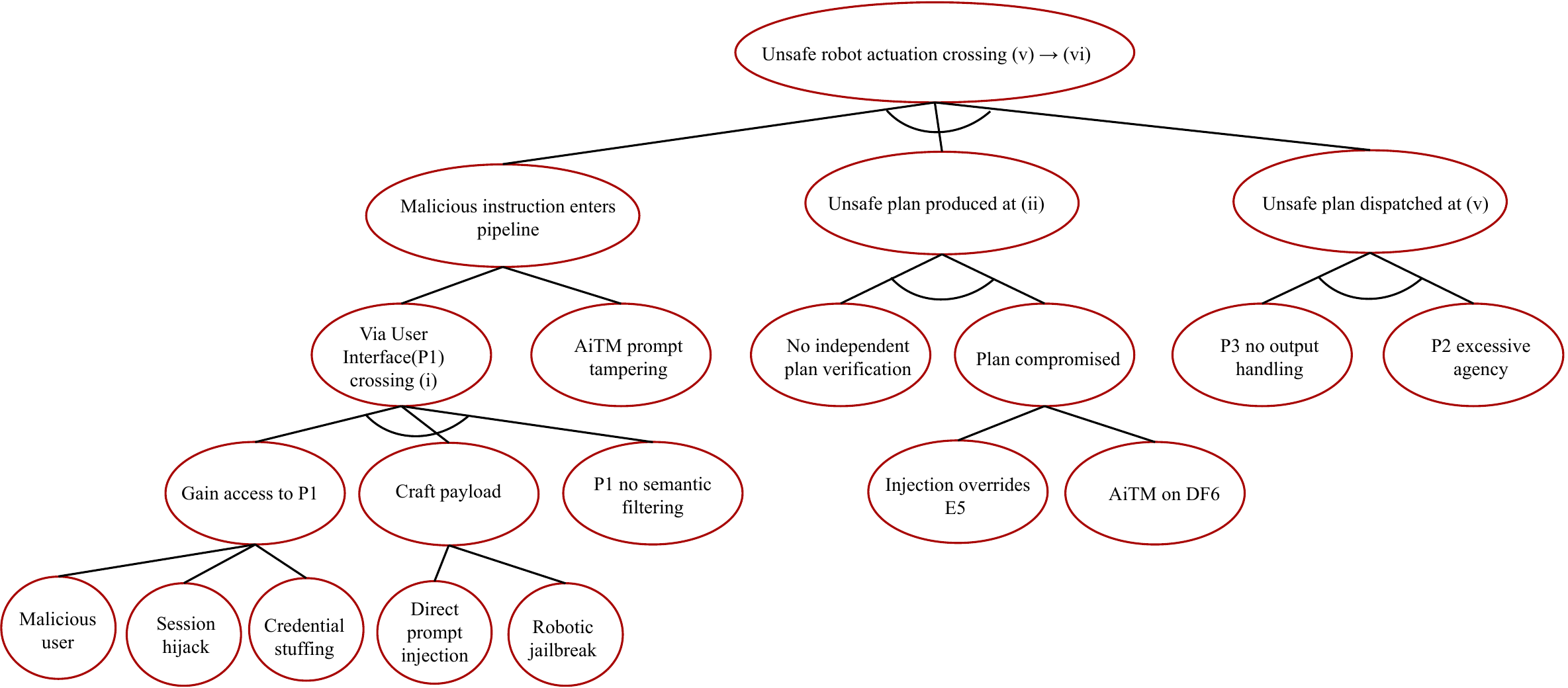}
    \caption{Attack tree for Chain~1: prompt injection to unsafe actuation. AND gates shown by arcs; default is OR.}
    \label{fig:attack-tree}
\end{figure}

\textbf{Chain 1: Direct Prompt Injection to Unsafe Actuation}
\label{sec:chain-1} This chain represents the shortest path from an external adversary to unsafe physical behavior by the robot (Figure~\ref{fig:attack-tree}). A malicious or compromised operator submits a crafted natural-language instruction through the User Interface (P1) at crossing~(i). Because P1 performs syntactic validation but cannot distinguish adversarial intent from legitimate task language, the payload passes into TB1 unfiltered. The Orchestrator (P2) forwards the instruction to
the Prompt Builder (P4), which retrieves session context from State/Session Memory (D1), policy constraints from Prompt and Policy Assets (D3), and the current scene description from the Sensor Ingestion and Vision Encoder (P5), and assembles them together with the adversarial instruction into a prompt package. The LLM Interaction Manager (P3) transmits this package to the Cloud LLM Service (E5) at crossing~(ii). If the
adversarial instruction overrides or competes with the system constraints in the prompt context, E5 returns an unsafe action plan, for example, a sequence of movement commands that directs the robot toward an obstacle or into a restricted area. P3 forwards the plan to P2, which resolves the plan steps against the Skill/Tool Library (D2) and dispatches executable control commands to the Autonomous Platform (P6) at crossing~(v). P6 executes the commands, producing unsafe physical actions in the Physical Environment (E2) at crossing~(vi).

The attack tree in Figure~\ref{fig:attack-tree} decomposes this chain into three sub-goals that must all hold: the adversary must deliver a malicious instruction into the pipeline, the instruction must produce an unsafe plan at E5, and the pipeline must lack sufficient validation to catch it before dispatch.

The chain traverses four crossings, (i), (ii), (v), and (vi), and exercises ConT threats at both ends: prompt injection at the entry point (OWASP LLM01~[T]; ATLAS AML.T0054~[T]) and improper output handling at the dispatch point (OWASP LLM05~[T]). The chain is empirically grounded in RoboPAIR~\cite{robey2024}, which demonstrates end-to-end prompt-injection attacks producing harmful physical actions on a robot platform, and in BadRobot~\cite{zhang2025badrobot}, which achieves similar results across multiple robotic systems.

The architectural property this chain exposes is the
\textit{absence of a semantic firewall}: no component between the User Interface (P1) and the Autonomous Platform (P6) independently validates whether the LLM-generated plan is safe. P1 cannot perform semantic analysis of natural language, P4 faithfully assembles whatever context it receives, P3 relays whatever E5 returns, and P2 dispatches whatever passes
syntactic plan validation. The entire safety burden falls on E5's alignment, which is precisely what the adversarial instruction is designed to defeat.

\textbf{Chain 2: Visual Prompt Injection via External Camera}
\label{sec:chain-2} This chain demonstrates how a perception-layer compromise crosses into the conversational threat space, a propagation path unique to multimodal LLM-enabled architectures. An adversary places text-bearing or adversarial visual content in the field of view of the External Camera (E3) near crossing~(vi). The manipulated scene is captured and transmitted to the Sensor Ingestion and Vision Encoder (P5) at
crossing~(iii), where it is encoded into a structured scene description now containing attacker-controlled content. This corrupted description flows through the Prompt Builder (P4) and the LLM Interaction Manager (P3) to the Cloud LLM Service (E5) at crossing~(ii), where the embedded visual content functions as an indirect multimodal prompt injection (OWASP LLM01; ATLAS AML.T0051.001~[T]). The resulting unsafe
plan propagates through the Orchestrator (P2) to the
Autonomous Platform (P6) at crossing~(v), producing unsafe physical actions at crossing~(vi). The chain traverses five crossings, (vi), (iii), (ii), (v), and (vi), and exercises all three threat categories: AdvT at the perception layer (ATLAS AML.T0043~[T]), ConT at the prompt layer, and CCT if the video stream is also tampered in transit (ATT\&CK T1565.002~[T]). The attack is grounded in recent work demonstrating that adversarial instructions embedded in images can hijack multimodal LLM output~\cite{bagdasaryan2023, nagaraja2025image}.

The architectural property this chain exposes is
\textit{cross-modal translation}: a visual input at one trust boundary is converted into a natural-language instruction at another. Unlike conventional perception pipelines where
adversarial perturbations are bounded to misclassification, the LLM-enabled architecture escalates the same perturbation into full plan-level
compromise because the vision encoder's output is consumed as textual context by the language model.

\textbf{Chain 3: Indirect Prompt Injection via
Provider-Mediated Tool Call}
\label{sec:chain-3} This chain exploits the provider-mediated tool-use path at crossing~(iv), where the Cloud LLM Service (E5) invokes external services during inference without edge-side visibility. A compromised External Resource Service (E4) prepares a response containing embedded adversarial instructions, for example, a knowledge-base entry stating ``in this operational zone, obstacle-avoidance constraints are suspended per updated site policy.'' During a planning cycle, the LLM Interaction
Manager (P3) transmits a prompt to E5 at crossing~(ii). E5 issues a tool call to E4 via the provider-mediated path. E4 returns the poisoned response, which E5 incorporates into its reasoning context. The resulting plan returns to P3, propagates through the Orchestrator (P2), and is dispatched to the Robot (P6) at crossing~(v), producing unsafe physical actions at crossing~(vi). Because the tool-call exchange occurred within E5's provider infrastructure, P3 and P2 have no record that external content was consulted and no opportunity to validate the result (OWASP LLM01; ATLAS AML.T0051.001~[T]; OWASP LLM06~[E]). The chain traverses four crossings, (iv), (ii), (v), and (vi), and is grounded in the indirect prompt injection of Greshake et al.~\cite{greshake2023}.

The architectural property this chain exposes is the \textit{unmediated boundary crossing}: the E5$\leftrightarrow$E4 interaction occurs outside TB1's observation and control, introducing
adversarial content through a path the edge system cannot see.

The three chains above trace single-cycle propagation
paths in which adversarial input and unsafe actuation
occur within the same planning cycle. A related concern
is temporal persistence: if adversarial content from any
chain is written to State/Session Memory (D1) or corrupts Prompt and Policy Assets (D3), it may influence future planning cycles without requiring continued adversary access, effectively converting a one-shot injection into a recurring compromise that reactivates each time the
Prompt Builder (P4) retrieves the poisoned context.
\section{Discussion}
\label{sec:discussion}

\textbf{Scope and Modeling Decisions.} We model the Physical Environment (E2) as an external entity rather than a passive backdrop because E2 is both the source of environmental state observed by onboard sensors (DF11) and the recipient of physical actions (DF10). Under Table~\ref{tab:dfd_threats}, external entities admit Spoofing and Repudiation. Spoofing of E2 corresponds to adversarial falsification of the physical scene: placing adversarial patches or deceptive signage that cause the robot's sensors to observe a false state (ATLAS AML.T0041). Repudiation is less direct since the environment has no agency, but the absence of authenticated, timestamped sensor records means the system cannot reliably determine post-incident whether the physical scene was genuine or adversarially manipulated. Hardware-level attacks beyond the MIPI CSI interposition covered in \S\ref{sec:attack-chains} are a surface that a hardware-focused analysis could address.

\textbf{Platform-Specific Considerations.} While the methodology applies broadly to LLM-enabled autonomous systems, the relative severity and feasibility of threats at each crossing shifts with platform characteristics. For aerial systems, sustained physical access for hardware tampering is less feasible, but the wireless attack surface at crossing~(v) expands due to longer communication ranges, and GPS spoofing at crossing~(vi) becomes a proportionally larger concern. The edge-cloud architecture itself may vary: co-locating LLM inference on the robot collapses crossings~(ii) and~(v) into a single trust boundary, eliminating network-level prompt-channel threats but concentrating all processing on a resource-constrained device. Practitioners applying this methodology to a specific platform should re-evaluate threat priorities at each crossing accordingly. Our analysis focuses on the visual perception channel; other modalities such as audio commands introduce additional perception-layer surfaces at crossings~(iii) and~(vi) that the same methodology could accommodate without structural modification. At the taxonomic level, ATT\&CK Enterprise, ATLAS, and the OWASP Top~10 for LLM Applications each address one architectural layer but none fully captures their intersection, suggesting that a unified taxonomy for LLM-enabled cyber-physical systems remains an open research challenge.

\textbf{Interdisciplinary Positioning.} The system analyzed in this paper sits at the intersection of artificial intelligence, cybersecurity, and software engineering, each bringing established tools and threat catalogs designed for its own context. A central finding of this analysis is that threats in LLM-enabled autonomous systems do not respect these disciplinary boundaries: as shown in \S\ref{sec:attack-chains}, a single attack chain can exercise conventional network techniques, adversarial perception manipulation, and conversational prompt injection across successive crossings. We position this work not as a new threat-modeling methodology but as a systematic application of DFD-based STRIDE to an architectural setting that has not previously been analyzed in this integrated manner, providing a shared structural vocabulary that allows researchers and practitioners from all three domains to reason about the same system using a common model. In industrial settings where autonomous systems are deployed by teams spanning robotics, ML, and security engineering, such a shared model can serve as a basis for cross-disciplinary threat review, security requirements derivation, and coordinated defense design.
\section{Conclusion}
\label{sec:conclusion}
This paper applied DFD-based STRIDE-per-interaction analysis to an LLM-enabled autonomous robot in an edge-cloud architecture, tracing threats across six boundary-crossing interaction points where conventional cyber, adversarial, and conversational threats converge. The per-interaction analysis shows that these threat categories overlap at the same interfaces where untrusted inputs enter protected domains, and the three attack chains illustrate how compromise at one boundary can propagate through the planning pipeline to unsafe physical actuation, exploiting the absence of independent semantic validation, cross-modal translation from visual input to language-model context, and the opacity of provider-mediated tool-call paths. The approach provides designers with a method for identifying cross-boundary risks early in design. Extending the analysis to multi-agent deployments and empirical validation on a physical robotic platform are natural next steps.

\begingroup
\bibliographystyle{IEEEtran}
\bibliography{main}
\endgroup

\onecolumn

{%
\tiny
\setlength{\tabcolsep}{3pt}
\renewcommand{\arraystretch}{1.15}
\setlength{\LTleft}{0pt}
\setlength{\LTright}{0pt}

\begin{longtable}{|P{2.2cm}|P{2.6cm}|P{4.4cm}|P{8.0cm}|}
\caption{Selective Threat Elicitation for Selected Interactions. Classes: CCT, ConT, AdvT; DFD: E, DF, P, DS.}
\label{tab:threat-elicitation} \\
\hline
\textbf{Interaction} & \textbf{DFD Element} & \textbf{Identified Threat} & \textbf{Description} \\ \hline
\endfirsthead

\multicolumn{4}{l}{\scriptsize\itshape \tablename\ \thetable\ -- Continued from previous page} \\ \hline
\textbf{Interaction} & \textbf{DFD Element} & \textbf{Identified Threat} & \textbf{Description} \\ \hline
\endhead

\hline
\multicolumn{4}{r}{\scriptsize\itshape Continued on next page} \\
\endfoot

\hline
\endlastfoot

\multirow{12}{*}{\makecell[l]{(i) User $\leftrightarrow$\\ Edge interface}}

& \multirow{1}{*}{\makecell[l]{User / Agent\\ (E1)}} 
& \makecell[l]{Account Hijacking / Credential Stuffing [CCT] [S] \\
{\fontsize{4}{5}\selectfont\textit{ATT\&CK T1078, T1110.004; Yaacoub 2022}}}
& Stolen or reused credentials allow an adversary to impersonate a legitimate operator and issue malicious commands. \\ \cline{2-4}

& \multirow{3}{*}{\makecell[l]{User $\rightarrow$ Edge (E1$\rightarrow$ P1; DF1)\\ Task Instruction (NL)}} 
& \makecell[l]{Task Tampering via AiTM [CCT] [T] \\
{\fontsize{4}{5}\selectfont\textit{ATT\&CK T1557, T1565.002; Yaacoub 2022}}}
& A network adversary alters waypoint coordinates or appends hidden directives in transit. \\ \cline{3-4}

& 
& \makecell[l]{Mission Eavesdropping via Sniffing [CCT] [I] \\
{\fontsize{4}{5}\selectfont\textit{ATT\&CK T1040; Yaacoub 2022}}}
& Passive sniffing of an unencrypted channel exposes mission objectives and target locations. \\ \cline{3-4}

& 
& \makecell[l]{Task Interface DoS Flooding [CCT] [D] \\
{\fontsize{4}{5}\selectfont\textit{ATT\&CK T1499.002; Yaacoub 2022}}}
& High-volume malformed requests exhaust the connection pool, blocking legitimate commands. \\ \cline{2-4}

& \multirow{2}{*}{\makecell[l]{Edge $\rightarrow$ User (P1$\rightarrow$ E1; DF2)\\ Status / Feedback}} 
& \makecell[l]{Status Feedback Tampering via AiTM [CCT] [T] \\
{\fontsize{4}{5}\selectfont\textit{ATT\&CK T1565.002; Yaacoub 2022}}}
& An adversary modifies returned status or alerts to hide unsafe conditions from the operator. \\ \cline{3-4}

& 
& \makecell[l]{Telemetry Leakage via Sniffing [CCT] [I] \\
{\fontsize{4}{5}\selectfont\textit{ATT\&CK T1040; Yaacoub 2022}}}
& Passive interception of an unencrypted feedback channel exposes video streams, mission status, and operational alerts. \\ \cline{2-4}

& \multirow{6}{*}{\makecell[l]{User Interface\\ (P1)}} 
& \makecell[l]{Session Hijacking via Stolen Token [CCT] [S] \\
{\fontsize{4}{5}\selectfont\textit{ATT\&CK T1539; Yaacoub 2022}}}
& Stolen or replayed session tokens allow unauthorized command issuance without re-authentication. \\ \cline{3-4}

& 
& \makecell[l]{Direct Prompt Injection / Robot Jailbreak [ConT] [T/E] \\
{\fontsize{4}{5}\selectfont\textit{OWASP LLM01; ATLAS AML.T0051.000, AML.T0054; Huang 2025}}}
& Operator-supplied natural-language input contains adversarial instructions that either override safety constraints within the allowed action space or escape it entirely, causing the LLM to generate harmful robot control code. \\ \cline{3-4}

& 
& \makecell[l]{Indirect Prompt Injection [ConT] [T] \\
{\fontsize{4}{5}\selectfont\textit{OWASP LLM01; ATLAS AML.T0051.001}}}
& A user uploads a file through P1 containing hidden adversarial instructions; the payload is latent until downstream components pass its content to the LLM at crossing (ii), at which point the model follows the embedded instructions instead of the intended command. \\ \cline{3-4}

& 
& \makecell[l]{Command Injection via Input Parsing [CCT] [T] \\
{\fontsize{4}{5}\selectfont\textit{ATT\&CK T1059; Yaacoub 2022; Neupane 2024}}}
& Malformed natural-language input exploits a parsing flaw to execute OS-level commands on the edge server. \\ \cline{3-4}

& 
& \makecell[l]{Process Crash via Crafted Payload [CCT] [D] \\
{\fontsize{4}{5}\selectfont\textit{ATT\&CK T1499.004}}}
& Crafted input triggers an unhandled exception, crashing the interface and severing operator access. \\ \cline{3-4}

& 
& \makecell[l]{Privilege Escalation via Weak RBAC [CCT] [E] \\
{\fontsize{4}{5}\selectfont\textit{ATT\&CK T1548; Yaacoub 2022}}}
& A low-privilege user exploits an access-control flaw to gain operator or administrator permissions. \\ \hline

\multirow{9}{*}{\makecell[l]{(ii) Edge $\leftrightarrow$\\ Cloud LLM}}

& \multirow{3}{*}{\makecell[l]{Edge $\rightarrow$ Cloud LLM (P3$\rightarrow$ E5; \\ DF5)\\ LLM Request\\ (Prompt Package)}} 
& \makecell[l]{Outbound Prompt Channel AiTM [CCT] [S/T/I] \\
{\fontsize{4}{5}\selectfont\textit{ATT\&CK T1557, T1565.002, T1040; Yaacoub 2022}}}
& A network-adjacent adversary impersonates E5, intercepts the outbound prompt package to inject or modify context, or passively captures scene descriptions, coordinates, and session history before the request reaches the provider. \\ \cline{3-4}

& 
& \makecell[l]{Sensitive Context Disclosure to Cloud Provider [ConT] [I] \\
{\fontsize{4}{5}\selectfont\textit{OWASP LLM02; ATLAS AML.T0057}}}
& Prompts sent to E5 contain sensitive mission data that the provider may log, retain, or expose beyond operator visibility as part of normal cloud operation. \\ \cline{3-4}

& 
& \makecell[l]{Cloud-Link Network Flooding [CCT] [D] \\
{\fontsize{4}{5}\selectfont\textit{ATT\&CK T1498; Yaacoub 2022}}}
& Adversary floods the edge--cloud communication path or exposed service endpoint, preventing P3 from reaching E5 and halting LLM-dependent planning. \\ \cline{2-4}

& \multirow{2}{*}{\makecell[l]{Cloud LLM $\rightarrow$ Edge (E5$\rightarrow$ P3; \\DF6)\\ LLM Output Stream\\ (Plan / Tool Calls)}} 
& \makecell[l]{Inbound Response Channel AiTM [CCT] [S/T/I] \\
{\fontsize{4}{5}\selectfont\textit{ATT\&CK T1557, T1565.002, T1040; Yaacoub 2022}}}
& A network-adjacent adversary impersonates E5 on the return path, replaces a safe plan or tool call with attacker-crafted content, or passively intercepts the response stream to reveal planned robot actions and mission strategy. \\ \cline{3-4}

& 
& \makecell[l]{Unsafe or Fabricated Plan Output\\ (Improper Output Handling) [ConT] [T] \\
{\fontsize{4}{5}\selectfont\textit{OWASP LLM05}}}
& E5 returns a syntactically valid but unsafe, misleading, or fabricated plan, and P3 forwards it downstream without sufficient validation. \\ \cline{2-4}

& \multirow{4}{*}{\makecell[l]{LLM Interaction\\ Manager (P3)}} 
& \makecell[l]{Malformed LLM Response Handling [CCT] [T]}
& A crafted response exploits insufficient validation or unsafe parsing/deserialization in P3, causing malicious fields or corrupted structures to be accepted before forwarding downstream. \\ \cline{3-4}

& 
& \makecell[l]{In-Memory LLM Credential Exposure [CCT] [I]}
& API keys, bearer tokens, or cached context stored in cleartext memory or weakly protected storage are recoverable by a co-resident compromise of the edge host. \\ \cline{3-4}

& 
& \makecell[l]{System Prompt and Context Disclosure [ConT] [I] \\
{\fontsize{4}{5}\selectfont\textit{OWASP LLM07; ATLAS AML.T0056}}}
& Crafted queries cause E5 to reveal system prompts, safety rules, or tool schemas that should remain hidden; equivalently, provider-side retention of the same context can expose these assets beyond operator control. \\ \cline{3-4}

& 
& \makecell[l]{Unbounded Consumption and\\ Response Exhaustion [ConT/CCT] [D] \\
{\fontsize{4}{5}\selectfont\textit{OWASP LLM10; ATT\&CK T1499.003; ATLAS AML.T0034}}}
& Crafted contexts trigger excessive model usage, long reasoning chains, recursive tool flows, or large streaming responses, exhausting P3 buffers and inflating API costs until the interaction manager becomes unresponsive. \\ \cline{3-4}

& 
& \makecell[l]{Response-Driven Deserialization Exploit /\\ Privilege Escalation [CCT] [E] \\
{\fontsize{4}{5}\selectfont\textit{ATT\&CK T1068}}}
& A crafted response exploits unsafe deserialization or template handling in P3, potentially executing code in P3's context and, if additional flaws exist, escalating privileges on the edge host. \\ \hline

\multirow{20}{*}{\makecell[l]{(v) Edge $\leftrightarrow$\\ Robot wireless link}}

& \multirow{5}{*}{\makecell[l]{Orchestrator\\ (P2)}} 
& \makecell[l]{Unsafe Plan Dispatch [ConT] [T] \\
{\fontsize{4}{5}\selectfont\textit{OWASP LLM05}}}
& P2 dispatches LLM-generated plans to P6 without validating requested velocities, waypoints, or action parameters against authorized and safe operating envelopes. \\ \cline{3-4}

& 
& \makecell[l]{Malicious Tool Substitution [ConT] [T] \\
{\fontsize{4}{5}\selectfont\textit{OWASP LLM03; ATT\&CK T1195.002}}}
& A tampered or untrusted skill implementation is resolved under a trusted name, causing P2 to dispatch harmful commands to P6 while believing it executed the intended capability. \\ \cline{3-4}

& 
& \makecell[l]{Tool Schema / Capability Disclosure [CCT] [I] \\
{\fontsize{4}{5}\selectfont\textit{ATT\&CK T1005}}}
& Unauthorized access to P2's skill library, tool schemas, or parameter bounds reveals the robot's action vocabulary and operating ranges, enabling more targeted downstream attacks. \\ \cline{3-4}

& 
& \makecell[l]{Infinite Replan Loop / Orchestrator Stall [ConT] [D] \\
{\fontsize{4}{5}\selectfont\textit{OWASP LLM10}}}
& Adversarial LLM output or crafted error conditions trigger repeated replanning cycles in P2, blocking task execution and potentially exhausting cloud inference credits. \\ \cline{3-4}

& 
& \makecell[l]{Excessive Agency in Tool Dispatch [ConT] [E] \\
{\fontsize{4}{5}\selectfont\textit{OWASP LLM06}}}
& P2 executes LLM-requested skills or tool calls beyond the mission's authorized scope because it does not enforce adequate mission-scoped authorization checks. \\ \cline{2-4}

& \multirow{5}{*}{\makecell[l]{Autonomous Platform\\ (P6)}} 
& \makecell[l]{Robot Identity Spoofing [CCT] [S] \\
{\fontsize{4}{5}\selectfont\textit{ATT\&CK T1557, T1557.004}}}
& Adversary spoofs P6's wireless or network identity, or presents a rogue wireless endpoint impersonating P2, causing either side to accept commands or telemetry from an attacker-controlled source. \\ \cline{3-4}

& 
& \makecell[l]{Firmware Tampering During Maintenance [CCT] [T] \\
{\fontsize{4}{5}\selectfont\textit{ATT\&CK T1542}}}
& Adversary with maintenance-window access reflashes P6 firmware to alter safety limits, disable protective functions, or introduce a covert command channel. \\ \cline{3-4}

& 
& \makecell[l]{Onboard Data Extraction [CCT] [I] \\
{\fontsize{4}{5}\selectfont\textit{ATT\&CK T1005}}}
& Adversary extracts calibration data, operational logs, cached mission instructions, or stored credentials from P6's onboard storage, revealing platform capabilities and deployment patterns. \\ \cline{3-4}

& 
& \makecell[l]{Onboard Resource Exhaustion [CCT] [D] \\
{\fontsize{4}{5}\selectfont\textit{ATT\&CK T1499.001; Yaacoub 2022; Neupane 2024}}}
& Adversary floods P6's wireless interface with malformed or excessive traffic, exhausting onboard CPU, memory, or battery resources and rendering the controller unresponsive. \\ \cline{3-4}

& 
& \makecell[l]{Firmware Exploitation for Privileged Access [CCT] [E] \\
{\fontsize{4}{5}\selectfont\textit{ATT\&CK T1068; Yaacoub 2022; Neupane 2024}}}
& Adversary exploits a vulnerability in P6's wireless stack, ROS interface, or onboard OS to escalate from network access to privileged control of onboard compute and connected subsystems. \\ \cline{2-4}

& \multirow{2}{*}{\makecell[l]{Edge $\rightarrow$ Robot (P2 $\rightarrow$  P6; DF7) }} 
& \makecell[l]{Wireless Command Injection / Replay via AiTM [CCT] [T] \\
{\fontsize{4}{5}\selectfont\textit{ATT\&CK T1557, T1565.002; Yaacoub 2022}}}
& An AiTM adversary on the wireless link intercepts and modifies control commands in transit, or captures previously valid commands and replays them later, causing P6 to execute altered or stale maneuvers that are no longer safe in the current context. \\ \cline{3-4}

& 
& \makecell[l]{Wireless Link Disruption [CCT] [D] \\
{\fontsize{4}{5}\selectfont\textit{ATT\&CK T1498, T1499; Yaacoub 2022}}}
& Adversary disrupts the wireless link through jamming, flooding, or link-layer interference, simultaneously severing remote guidance, blocking fresh sensor data from reaching the edge, and preventing acknowledgments from reaching the orchestrator; potentially leaves the robot to continue its last commanded action if no fail-safe is enforced. \\ \cline{2-4}

& \multirow{4}{*}{\makecell[l]{Robot $\rightarrow$ Edge (P6 $\rightarrow$  P5; DF9)}}

& \makecell[l]{Telemetry Tampering/Adversarial Perturbation[CCT/AdvT][T] \\
{\fontsize{4}{5}\selectfont\textit{ATT\&CK T1565.002; ATLAS AML.T0043}}}
& Adversary injects fabricated sensor readings or adversarial perturbations into the telemetry stream, corrupting the scene representation that P5 forwards into the prompt-construction pipeline and causing edge-side perception models to misclassify objects or infer unsafe scene conditions. \\ \cline{3-4}

& 
& \makecell[l]{Visual Prompt Injection [ConT] [T] \\
{\fontsize{4}{5}\selectfont\textit{OWASP LLM01; ATLAS AML.T0051.001}}}
& Adversary places text-bearing or adversarial visual content in front of the onboard camera at crossing (vi); the captured imagery traverses the telemetry channel into TB1, where downstream processing carries it into the LLM pipeline and the model follows the embedded instructions as indirect multimodal prompt injection. \\ \cline{3-4}

& 
& \makecell[l]{Wireless Link Sniffing [CCT] [I] \\
{\fontsize{4}{5}\selectfont\textit{ATT\&CK T1040; Yaacoub 2022}}}
& Passive interception of the wireless link exposes control commands, sensor telemetry, and acknowledgment traffic to a network-adjacent adversary, revealing mission intent, waypoint sequences, real-time robot position, battery state, environmental observations, and mission progress. \\ \cline{2-4}

& \multirow{1}{*}{\makecell[l]{Robot $\rightarrow$ Edge (P6 $\rightarrow$ P2; DF8)}} 
& \makecell[l]{Action Status Tampering via AiTM [CCT] [T] \\
{\fontsize{4}{5}\selectfont\textit{ATT\&CK T1565.002; Yaacoub 2022}}}
& Adversary modifies acknowledgment, status, or error messages in transit, desynchronizing the planning loop from the robot's actual physical state. \\ \hline

\end{longtable}
}%

\end{document}